\documentclass[aps,prl,reprint,floatfix,superscriptaddress]{revtex4-2}

\usepackage{algorithm}
\usepackage{algpseudocode}
\usepackage{amsmath}
\usepackage{comment}
\usepackage{amsthm}
\usepackage{bbold}
\usepackage{graphicx}
\usepackage{braket}
\usepackage{xcolor}
\PassOptionsToPackage{hyphens}{url}

\usepackage[colorlinks=true, allcolors=blue]{hyperref}

\DeclareMathOperator{\tr}{Tr}
\newcommand\norm[1]{\lVert#1\rVert}
\newcommand{\LL}{\mathcal{L}}

\newcommand{\ii}{ {\rm i} }
\newcommand{\ave}[1]{{\langle #1\rangle}}
\usepackage{comment}

\begin{document}
\title{
Opening Krylov space to access all-time dynamics via dynamical symmetries
}
\author{Nicolas Loizeau}
\affiliation{Niels Bohr Institute, University of Copenhagen, Copenhagen, Denmark}
\author{Berislav Bu\v ca}
\affiliation{Niels Bohr Institute, University of Copenhagen, Copenhagen, Denmark}
\affiliation{Universit\'{e} Paris-Saclay, CNRS, LPTMS, 91405, Orsay, France}
\affiliation{Clarendon Laboratory, University of Oxford, Parks Road, Oxford OX1 3PU, United Kingdom} 
\author{Dries Sels}
\affiliation{Department of Physics, New York University, New York, NY, USA}
\affiliation{Center for Computational Quantum Physics, Flatiron Institute, New York, NY, USA}

\date{\today}

\begin{abstract}

Solving short and long time dynamics of closed quantum many-body systems is one of the main challenges of both atomic and condensed matter physics. For locally interacting closed systems, the dynamics of local observables can always be expanded into (pseudolocal) eigenmodes of the Liouvillian, so called dynamical symmetries. They come in two classes - \emph{transient} operators, which decay in time and \emph{perpetual} operators, which either oscillate forever or stay the same (conservation laws). These operators provide a full characterization of the dynamics of the system. Deriving these operators, apart from a very limited class of models, has not been possible.
Here, we present a method to numerically and analytically derive some of these dynamical symmetries in infinite closed systems by introducing a naturally emergent open boundary condition on the Krylov chain. This boundary condition defines a partitioning of the Krylov space into system and environment degrees of freedom, where non-local operators make up an effective bath for the local operators.
We demonstrate the practicality of the method on some numerical examples and derive analytical results in two idealized cases. Our approach lets us directly relate the operator growth hypothesis to thermalization and exponential decay of observables in chaotic systems and provides a powerful approach for computing notably challenging many-body dynamics.
\end{abstract}

\maketitle

{\it Introduction -- } In closed quantum many-body systems, generic observables thermalize under unitary dynamics, spreading over the entire available phase space \cite{DAlessio2016}. The dynamical behavior of real-world systems can however be vastly different, ranging from overdamped decay to almost perpetual motion. Examples of non-trivial dynamical behaviors in quantum mechanics include many-body localization \cite{Nandkishore2015, lbits1,Abanin2019, Smith2016, Alet2018,DaleyMBL}, time crystal \cite{Wilczek2012, TCReview,Else2016, Else2020, Buča2019, Sacha_2015,VedikaLazarides,Sacha,Passarelli_2022,MeasurementTC,carollo2024quantum,DiehlTC,HadisehMulti,Shammah,Fabrizio1,Booker_2020,PhysRevA.108.L041303,PhysRevA.105.L040202,Wu_2024,Subhajit,PhysRevA.106.022209}, many-body scars \cite{Turner2018, Moudgalya2018, PhysRevB.101.165139,Schecter2019,Serbyn2021,chandran2022quantum,embed,HoshoScars,BridgingScars,Tom1,Leonardo1,Leonardo2,Leonardo3,Leonardo4,quasisymmetry,Lei}, fragmentation \cite{ZanardiFRAG,Fragmentation1,Fragmentation2,fragmentationSanjay,LOCfrag2,LOCfrag3,strictlylocalfrag,nicolau2023local,Zhang_2023,Sid,SanjayNew,Pozsgay,PhysRevB.107.205112,ArnabFragmentation,openFRAG}, and related forms of ergodicity breaking (e.g. \cite{Kormos_2016,Majidy_2024,CatQUBIT,Henrik,Olalla1,Olalla2,Kollath1,ETH1,PhysRevLett.134.073604,Buca2022Syn,Hosho,Jad4,Claeys2022emergentquantum,FiniteFreqDrude,AndreasStark1,PhysRevResearch.7.013178,PhysRevA.105.L020401,Shovan,PhysRevLett.132.020401}). 
Stationary and non-stationary quantum dynamics can respectively be studied in terms of conserved quantities and dynamical symmetries. For a given Hamiltonian $H$, a dynamical symmetry is an observable that satisfies $[H, A_\omega]=-\omega A_\omega$ \cite{Buča2019,Buca2020, Marko2,Tindall2020, Buca2023,PhysRevLett.132.196601,Chinzei}.
In principle, the dynamics of any quantum system can be expanded in the basis of the dynamical symmetries of $H$. For example, when time evolving an observable $O$ with initial state $\rho$ we have $\tr(O(t)\rho)=\sum_\omega e^{i\omega t} \mu_\omega$ where $\mu_\omega = \tr(A_\omega \rho) \tr(A^\dagger_\omega O(0))$. Thus, knowing the dynamical symmetries is crucial as it provides direct access to the dynamics of any operator and initial state.

These dynamical symmetries and conserved quantities are eigenmodes of the Liouvillian $\LL = [H, \cdot]$ (not to be confused with the related Lindbladian \cite{breuer2002theory}) therefore it is natural to work in Krylov space, where the Liouvillian is already tri-diagonal. In general, the frequency $\omega$ of a dynamical symmetry can be complex, and the real part is responsible for oscillatory behavior while the imaginary part is responsible for exponential decay in time (or growth). Such decaying dynamical symmetries are referred to as transient and the ones with purely real frequencies (including $0$) are referred to as \emph{perpetual} dynamical symmetries (including conservation laws).

Crucially, for locally interacting systems and initial states with short-range correlations, the only relevant dynamical symmetries are those satisfying the property of pseudolocality \cite{Tomaz_Prosen_1998,Doyon,Doyon2,Doyon3,Buca2023}. Note that sometimes, these dynamical symmetries are referred to as a spectrum generating algebra (SGA, e.g. \cite{SGA,scarsdynsym2,Serbyn2021,chandran2022quantum}), however, we emphasize that dynamical symmetries do not form an algebra in general, and conversely SGAs are not necessarily pseudolocal and hence might not affect the dynamics.

Quantum mechanics is unitary, so in finite closed systems, however, the spectrum of the Liouvillian is purely real. This raises the question: How can we probe the thermodynamic limit properties of a system by studying the spectral properties of the Liouvillian? Similar approaches have been first developed to compute time dependent correlation functions and are referred to as pseudomode expansions~\cite{Mori1965, Mori2024, Dalton2001, Teretenkov2024b,Teretenkov2024, Uskov2024}. 
In particular, Refs \cite{Teretenkov2024, Uskov2024} suggest adding dissipation in the Krylov basis and taking the limit of vanishing dissipation or by studying the Ruelle-Pollicott resonances of an effective Liouvillain \cite{prosen2002ruelle,RP1,RP2,RP3,RP4,RP5,RP6}. Both of these approaches allow for access to the dynamical symmetries. However, the studies so far have been limited to close to equilibrium dynamics. Equally importantly, the (pseudo)locality of the dynamical symmetries has not been previously investigated, even though it is precisely the pseudolocal operators that affect the non-equilibrium physics.

We propose a more rigorous approach that consists in truncating the Krylov chain by introducing an open boundary condition that leads to a decomposition into system and environment at the level of the Krylov space. Local quantities are supported in the system while nonlocal quantities are part of the environment. This allows for understanding which dynamical symmetries are pseudolocal and hence relevant. 

This framework is practical provided that one is interested in eventually measuring local quantities and provided that some smoothness condition holds on the operator representation in Krylov space.

{\it Method -- } 
We will assume that the initial state has exponential clustering of connected correlation functions of local operators. We begin by recalling the concept of eternal equilibrium \cite{Buca2023}. A locally interacting system when quenched from a state that has clustering (i.e. short range correlation), provided that certain assumptions hold - normality of the state at all times, short range correlations, etc. (see \cite{Buca2023} for details), is in an equilibrium-like state, 
\begin{equation}
\rho(t)=\frac{1}{Z} \exp{\bigg(\sum_u \mu_u e^{\ii \lambda_u t} A_u\bigg)}, \label{tgge}
\end{equation}
where $\mu_u$ are the chemical potentials set by the initial state, and $\lambda_u$ are the possibly complex eigenfrequencies of the infinite Hamiltonian, $A_u$ are the pseudolocal dynamical symmetries, and $Z$ is the normalization constant. Pseudolocality means that it satisfies a generalized requirement of extensitivity, $\ave{(A)^\dagger_u A_u}_{cc} \propto N$ and existence of overlap with an extensive observable that is translationally invariant $O:=\sum_x\tau_x(o)$, $\exists\lim_{N \to \infty} \frac{1}{N}\ave{OA_u}_{cc}, \forall O$ in the thermodynamic limit, where $\ave{CD}_{cc}=\ave{CD}_0-\ave{C}_0\ave{D}_0$, and $\ave{C}_0=\tr\left(\rho(0)C\right)$ and we used translational invariance. This connected correlator allows us to define a suitable inner product that we will use later \cite{Buca2023}. Intuitively, \eqref{tgge} can be understood as the result of a quench from the thermal state of $O_0$, $\rho(0)=e^{-\beta O_0}/\tr(e^{- \beta O_0})$ to Hamiltonian $H$. We have $\rho(t)=e^{-\beta O_0(t)}/\tr(e^{- \beta O_0(t)})$ where $O_0(t)=e^{-iHt}O_0 e^{iHt}$ and $O_0(t)$ can be decomposed into the basis of dynamical symmetries.

In order to construct the dynamical symmetries we will utilize a novel Krylov space approach. We first recall how to construct the Krylov space of an observable $O_0$ under Hamiltonian $H$ using Lanczos algorithm.
The algorithm starts with a seed operator $O_0$ then constructs an orthonormal basis of operators by recursively applying the Liouvillian $\mathcal{L}$ to $O_0$ while orthonormalizing at each step \cite{Parker2019, Nandy2024,Claeys,PhysRevB.111.054302}.
The first iteration is given by $O_1 =\LL O_0/b_1 = [H,O_0]/b_1$ and
$b_1=\norm{\LL O_0}$
and for $n>2$,
\begin{align}
    O_{n}' & = \LL O_{n-1}-b_{n-1}O_{n-2}, \nonumber \\
    O_n & = \frac{O_{n}'}{b_{n}}, \nonumber \\
    b_n &= \norm{O_n' }.
    \label{eq:lanczos}
\end{align}
where  $\norm{O}^2 = \frac{1}{2^N}{\rm Tr}[O^2]$ \cite{footnote}.
The algorithm yields an orthonormal `Krylov-basis' $\{O_n\}$ and `Lanczos coefficients' ${b_n}$.

The power of the method is that time evolution in the Krylov basis can be mapped to a 1D single particle problem.
Consider $O(t)$ \cite{footnote1} an operator evolving in the Heisenberg picture. If we expand it in the Krylov basis : $O(t)=\frac{1}{2^N}\sum_n i^n \varphi_n(t) O_n$ then the coefficients $\varphi_n(t)$ evolve like
\begin{align}
    \partial_t \varphi_n=b_{n}\varphi_{n-1}-b_{n+1}\varphi_{n+1}, \quad \varphi_n(0)=\delta_{n0}.
    \label{eq:chain}
\end{align}
In other words, $\varphi$ can be interpreted as a single particle hopping on a 1D chain.
The method has gathered recent interests for probing operator complexity in quantum chaos \cite{Parker2019,Barbón2019,Dymarsky2020,Rabinovici2022,Trigueros2022,Takahashi2024,Bhattacharjee2022}, 
for probing for hydrodynamics \cite{Uskov2024, Wang2024, Stuart2024, paulistrings1}, or studying Floquet systems \cite{Mitra1, Mitra2, Mitra3, Mitra4, Kolganov2025} among other applications.
\begin{figure}[t]
\centering
\includegraphics[width=0.4\textwidth]{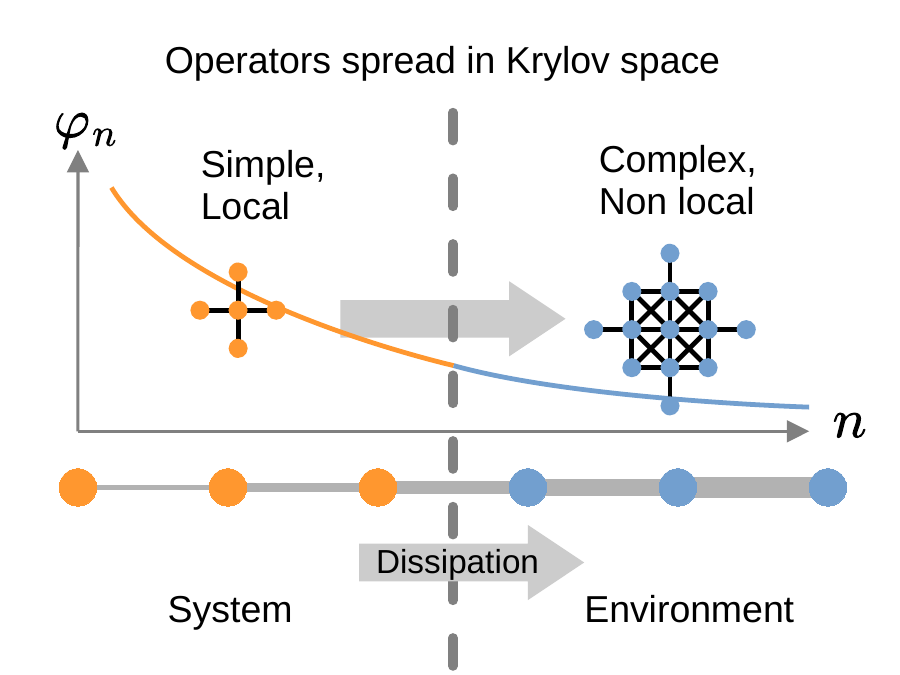}
\caption{In Krylov space, dynamics can be mapped to a single particle hoping on a 1D half chain. We split the chain into a system (left) where the hopping coefficients are exactly known and an environment (right). The left part of the chain corresponds to simple operators supported on $k$-local Pauli strings while the environment correspond to complex non-local operators. We wish to describe the dynamics solely based on the system degrees of freedom, but we need to allow for flow into the environment and back which leads to non-hermitian boundary conditions on the left part of the chain.}
\label{fig:drawing}
\end{figure}

Numerically, the crux of the method lies in being able to efficiently compute nested commutators. For local spin models, this can be done rather efficiently using Pauli strings representation. We use the Julia package \href{https://paulistrings.org/}{PauliStrings.jl} ~\cite{paulistrings1, paulistrings2}. 

In practice, we cannot compute all the Lanczos coefficients $b_n$. Computing $\LL O_n$ becomes increasingly difficult with $n$ as $O_n$ becomes increasingly non local . We would like to infer the spectral properties of $\LL$ in the thermodynamic limit from a finite, hopefully sufficiently large and informative set of Lanczos coefficients. To that extend we propose to split the Lanczos chain into a system and an environment, as shown in figure \ref{fig:drawing}. The system corresponds to the initial (left) part of the chain, where the hopping coefficients are exactly known. The hopping coefficients in the environment, which is the right part of the chain, are unknown. 

To truncate the 1D chain at site $L$, we assume that $\varphi_{n}$ is sufficiently smooth such that we can locally approximate it by a linear function around $n=L$ i.e. $\varphi_{L+1}\sim \varphi_{L}+(\varphi_{L}-\varphi_{L-1})$. This yields the dissipative boundary condition at site $L$:
\begin{align}
    \partial_t \varphi_L = (b_L+b_{L+1})\varphi_{L-1}-2b_{L+1}\varphi_{L}.
    \label{eq:psi_l}
\end{align}
In matrix form, the Liouvillian of the 1D truncated chain is therefore
\begin{equation}
\Tilde{\LL}=i\begin{pmatrix}
0   & -b_1   & 0     & \cdots & 0     & 0  \\
b_1   & 0   & -b_2   & \cdots & 0     & 0  \\
0     & b_2   & 0   & \cdots & 0     & 0  \\
\vdots& \vdots& \vdots& \ddots & \vdots\\
0     & 0     & 0     & \cdots & 0  & -b_L \\
0     & 0     & 0     & \cdots & b_L+b_{L+1} & -2b_{L+1}
\end{pmatrix}\label{eq:open_chain}
\end{equation}
Note that the dissipation makes the problem non hermitian. In contrast to the method used in refs.~\cite{Uskov2024, Teretenkov2024}, our method does not require any extrapolation of the Lanczos coefficients. Additionally, it does not introduce any extra parameters such as dissipation rate. 

In real space, the boundary condition can be interpreted as separating the local degrees of freedom from the nonlocal ones, and grouping the nonlocal DOF in a bath. In fact, the operators on the right side of the Krylov chain are dominated by k-local Pauli strings with $k>L$.

\begin{figure}
\centering
\includegraphics[width=0.49\textwidth]{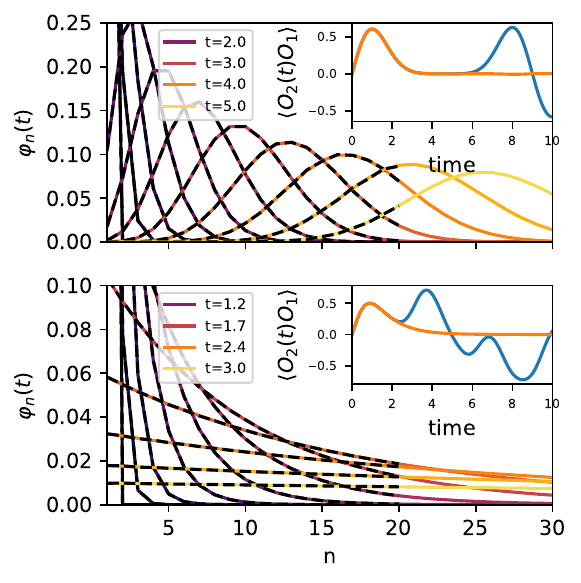}
\caption{Krylov chain dynamics for the ideal cases of $b_n=n$ (bottom) and $b_n=\sqrt n$ (top). We start with $\varphi_n = \delta_{0,n}$ and show $\varphi_n(t)$ for different $t$. Each color corresponds to a particular time, dark is early, yellow is late. Colors are exact results and dashed lines correspond to the open Krylov chain $\eqref{eq:open_chain}$. The insets show the time evolution of $\tr(O_1 O_0(t))$. The orange line is obtained with our newly proposed boundary condition while the blue line used standard Dirichlet boundary condition $\varphi_{21}=0$, in the latter case the wavepacket bounces of the edge of the Krylov chain, leading to unphysical revivals.}
\label{fig:ideal_chain}
\end{figure}

{\it Exact results -- }
Having defined our method, let's first examine some idealized cases before discussing our numerical results for actual spin chains. For generic chaotic Hamiltonians, the Lanczos coefficients $b_n$ are expected to grow linearly while in the integrable case, they commonly grow as $\sim \sqrt n$ \cite{Parker2019, Cao2021}. 
In figure \ref{fig:ideal_chain} we show how our modified boundary condition \eqref{eq:open_chain} affects the dynamics of the Krylov chain in these two ideal cases. Clearly, this boundary condition allows us to truncate the chain at some finite $n$, say $n=20$ as shown in Fig.\ref{fig:ideal_chain}, without significantly altering the dynamics, which remain indistinguishable from those of the infinite chain. In the absence of this boundary condition, i.e. using normal Dirichlet boundary conditions $\phi_{n+1}=0$, the wave function bounces back on the edge of the chain, leading to nonphysical revivals in the dynamics of the initial system.

\begin{figure}
\centering
\includegraphics[width=0.49\textwidth]{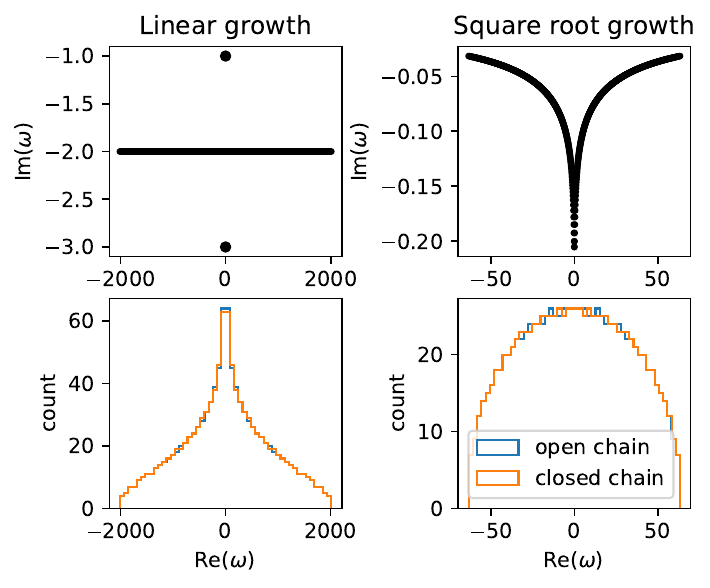}
\caption{\textbf{Top:} Liouvillian spectrum of the ideal case $b_n=n$ (left) and Liouvillian spectrum of the ideal case $b_n=\sqrt n$ (right).
\textbf{Bottom:} distribution of $\text{Re}(\omega)$ for the Liouvillian with truncated Krylov chain (orange) or open boundary condition as defined in eq \eqref{eq:open_chain} (blue). In the case of the simple truncated chain, $\LL$ is hermitian and its spectrum is real. When introducing the boundary condition, the spectrum becomes complex but the distribution of the real parts is almost identical. }
\label{fig:ideal_spectrum}
\end{figure}

This modification is clearly reflected in the Liouvillian spectrum of the problem, shown in figure \ref{fig:ideal_spectrum}. In both cases, the open boundary condition makes the Liouvillian spectrum complex, but without affecting the real part of the spectrum much. In particular, in the linear case, the spectrum is almost identical to the truncated chain one, but shifted by $-2\alpha i$ in the complex plane (with $\alpha$ the growth rate of the Lanczos coefficients).

When $b_n=n$ or $b_n=\sqrt n$, we can analytically express the eigenvectors of $\Tilde\LL$ in terms of Meixner and Hermite polynomials respectively, as shown in the supplements. Note that similar but different toy models with linearly growing Lanczos coefficients and dissipation are exactly solved in refs \cite{Parker2019,Teretenkov2024, Balasubramanian2022, Bhattacharjee2023} and some also exhibit Meixner polynomials. 

The linear case is particularly interesting because it lets us relate the operator growth hypothesis to thermalization. 
In this case, the eigenvalues $\omega$ are the roots of $P=\left(\omega+2i(L+1)\right)M_L(\omega)-(2L+1)lM_{L-1}(\omega)$
where $M_n$ is the $n$th Meixner polynomial \cite{oeis_m}. 

$P$ has two trivial roots $\omega=-i$ and $\omega=-3i$ with respective eigenvectors $\varphi_{n} \propto 1$ and $\varphi_{n} \propto 2n-1$. We empirically find that all other roots have $\textup{im}(\omega)=-2i$ (cf fig \ref{fig:ideal_spectrum}). This means that if the $b_n$ grow linearly with rate $\alpha$, then in this chaotic model, all observables time decay exponentially with rate $-2\alpha$. Although we do not prove that $\textup{im}(\omega)=-2i$ for all non-trivial roots, we numerically checked it up to $n=2^{12}$. In the next section we will see that this result is particularly insightful when studying real models like spin systems: the eigenmodes of chaotic models accumulate on the $-2i\alpha$ line.

{\it Spin chains results -- }
We will now present results for real spin systems. We use the framework developed in the previous section to address the following questions : Given a Hamiltonian $H$ and a density matrix $\rho$ that we quench from, or a operator $O_0$ that we want to time evolve in the Heisenberg picture, what kind of dynamics is the system subject to ? In Krylov space, this is equivalent to constructing the Krylov space of $O_0$ under $H$, constructing the tridiagonal dissipative Liouvillian \eqref{eq:open_chain} and studying its spectral properties.

\begin{figure}[h]
\centering
\includegraphics[width=0.49\textwidth]{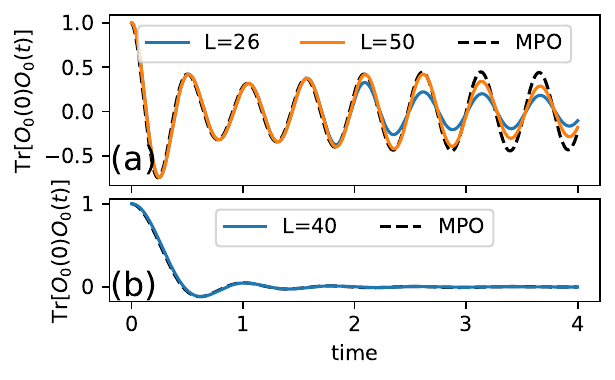}
\includegraphics[width=0.49\textwidth]{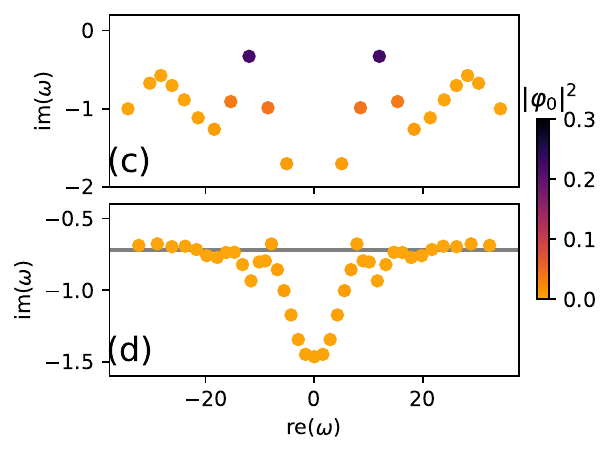}
\caption{
(\textbf{a, b}) Time evolution of the initial operator $O_0$ for the $XXZ$ model and chaotic chain respectively. In the XXZ model, any operator that has overlap with the dynamical symmetry will oscillate perpetually. In the chaotic model, all observables decay because there are no perpetual (real valued) dynamical symmetries. These results are obtained by solving for $\varphi_0(t)$ by diagonalizing the open Krylov chain \eqref{eq:open_chain}. For reference, the dashed lines show results for tensor networks simulations with $N=40$ spins. The orange line is obtained by truncating the operators in the Pauli string space in addition to truncating the Krylov chain.
(\textbf{c, d}) Spectrum of the open Krylov chain \eqref{eq:open_chain} for the $XXZ$ model and chaotic chain. In the XXZ model, we recover the dynamical symmetry at $\textup{re}(\omega)=12$ (black). In the chaotic chain, the grey horizontal line indicates $\textup{im}(\omega)=-2\lambda$ where $\lambda$ is the growth rate of the Lanczos coefficients. The color shows $\psi_0$, the first coefficient of the dynamical symmetry in the $O_n$ basis. Darker markers correspond to more local operators.  }
\label{fig:main}
\end{figure}

In figure \ref{fig:main} we show numerical results for two canonical models, i.e. the XXZ chain and the transverse field Ising chain. In particular, for the XXZ model, we take the Hamiltonian:
\begin{equation}
    H = \sum_i \left( s^x_i s^x_{i+1}+s^y_i s^y_{i+1}+\Delta s^z_i s^z_{i+1} + h s^z_i \right),
    \label{eq:XXZH}
\end{equation}
with $\Delta=-\frac{1}{2}$ and $h=2$. This model is integrable and  is known to exhibit a dynamical symmetry, the first order of which is $Q_3=\sum_i s_i^+s_{i+1}^+s_{i+2}^+ + s_i^-s_{i+1}^-s_{i+2}^-$ \cite{Medenjak2020, Zadnik2016}. We use this as the initial operator $O_0=Q_3$. We contrast this to a chaotic chain~\cite{Banuls} with Hamiltonian:
\begin{align}
\label{eq:CC}
    H = \sum_i \big(s^x_is^x_{i+1}-1.05 s^z_i +\frac{1}{2} s^x_i\big).
\end{align}  
Here we use
$O_0 = \sum_i \left(1.05 s^x_is^x_{i+1}+s^z_i\right)$
as the initial operator. This operator has no overlap with the Hamiltonian therefore the trivial conserved quantity $H$ will not be part of the Krylov subspace. We chose this model because it is a simple chaotic model whose Lanczos coefficients have previously been studied~\cite{ Parker2019, paulistrings1}. Note that both models and initial operators are translation-invariant. We numerically take advantage of this by only storing a unit cell instead of the extensive operators, as described in \cite{paulistrings1}.

The spectrum of the open Krylov chain \eqref{eq:open_chain} for the two example models is shown in \ref{fig:main}(c,d). Purely real eigenvalues correspond to oscillating modes, while eigenvalues with a negative imaginary part are decaying modes. In the XXZ model (fig. \ref{fig:main}c), we recover (to good approximation) the known dynamical symmetry at $\omega=12$. It is possible to significantly improve the results by truncating the operators in the Pauli strings representation during the Lanczos algorithm in addition to truncating the Krylov chain. These results are shown in orange on figure \ref{fig:main}a and the method is discussed in the supplements.

The persistent dynamical symmetry at $\omega=12$ leads to oscillatory dynamics of operators that have overlap with it (fig. \ref{fig:main}a). All other modes decay significantly faster. In the chaotic model (fig. \ref{fig:main}d), there are no perpetual (purely real valued) dynamical symmetries, all the modes are decaying with a decay rate centered around $\textup{im}(\omega)=-0.72i$ and the characteristic dynamics are transient (fig. \ref{fig:main}b). The latter can be directly related to the growth rate of the Lanczos coefficients as discussed in the previous section, thus substantiating the universality of the time decay for linearly growing Lanczos coefficients. 

For reference, we benchmark our results with matrix product operators (MPO) simulations (dashed line on figure \ref{fig:main}ab). At longer times, the Krylov results suffer errors that can be attributed to the non fully converged imaginary part of the dynamical symmetry.

In the Krylov basis, a dynamical symmetry can be expressed as $A_u=\sum_n \varphi_{u,n} O_n$. Note that if $O_0$ is $k$-local, then $O_n$ is at most $n+k$-local. Therefore in this representation, we can quickly visualize how local the extracted dynamical symmetries are. In the XXZ model (fig. \ref{fig:main} (c), we find that only the real-valued dynamical symmetry is local. In the chaotic model (fig. \ref{fig:main} (d)), all the modes are extremely non-local. This is of crucial importance when measuring local observables: even if the models may have long-time oscillating modes, these do not impact local measurements if they are non-local.

\begin{figure}
\centering
\includegraphics[width=0.49\textwidth]{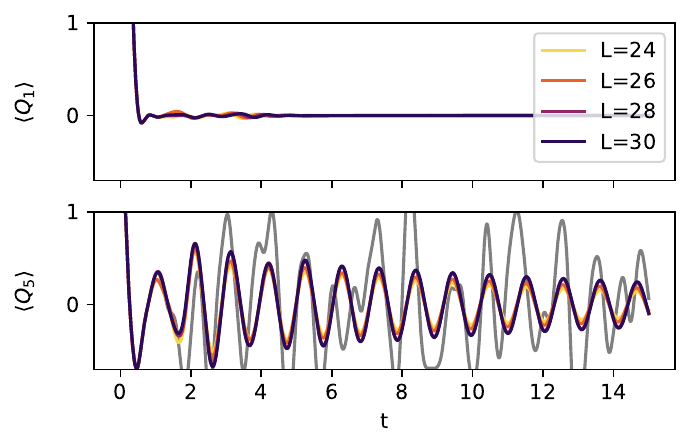}
\caption{Quench from $\ket{\psi_0}=2^{N/2}\left(\ket{0}+\ket{1}\right)^{\otimes N}$ in the XXZ model. $Q_5$ has overlap with the dynamical symmetry and therefore exhibit long time oscillations. These simulations where performed using our open Krylov chain technique. For comparison, the grey line in the second plot shows results for a truncated Krylov chain without using our boundary conditions. $L$ correspond to the length of the Krylov chain.}
\label{fig:quench}
\end{figure}

Although one can find the dynamical symmetries, it is still numerically hard to quench from a thermal state because of the estimate of $e^{-\beta O_0(t)}$ requires computing the exponential of a sum of non-commuting operators. Using a Pauli string representation for the operators, is however rather simple to quench from stabilizer states. An example of such a quench is shown in figure~\eqref{fig:quench}, where we time evolve $Q_1=\sum_i s_i^++s_i^-$ and $Q_5$ the 3rd order 5-local expansion of the dynamical symmetry in the XXZ chain, starting from the state $\ket{\psi_0}=2^{-N/2}\left(\ket{0}+\ket{1}\right)^{\otimes N}$ \cite{Medenjak2020}. 
Pure states cannot be efficiently represented in the Pauli string basis \cite{loizeau2023}, however it is straightforward to compute the expectation value $\bra{\psi_0}O(t)\ket{\psi_0}$ after time-evolving $O$ in Krylov space. Note that quenches from a similar state have also been very recently studied in Krylov space in ref \cite{shirokov2025} for the transverse Ising model.

{\it Discussion -- }
In this work, we have developed a technique for solving the dynamics of quantum many-body systems based on truncating the Krylov chain. Instead of naively truncating the chain, we introduce an open boundary condition that reflects the spreading of non-local operators into a bath and encodes continuity of the Krylov wave function's derivative. This boundary condition avoids retarded interactions and knowledge of the fine structure of the bath, but in order to do so assumes some smoothness of the Krylov representation of the operator. The latter is the main limitation of the method. Combined with Pauli strings approach \cite{paulistrings1, Angrisani2025, Dehaene2003}, this method represents an effective numerical technique to study the Liouvillian spectrum of infinite closed quantum systems.

In chaotic models, the effectiveness of the method is closely tied to the operator growth hypothesis \cite{Parker2019}. Indeed, chaotic models have linearly growing Lanczos coefficients, which ensure the validity of our boundary solution (smoothness of the $b_n$) and also force all the dynamical symmetries to decay fast in time. In other models, our method makes it clear how to compute short and long-time dynamics using the spectrum of the Liouvillian. This method might be useful in computing short and long lived dynamics in many-body systems, and thus settling one the long-standing open problems in atomic and condensed matter physics. In future work we plan to generalize and apply the method to quantum circuits \cite{Claeys} (where the Liouvillian is an upper Hessenberg matrix), other pseudolocal dynamical symmetries, such as semilocal ones \cite{LenartFolded1,Semilocal2,Semilocal1} and the pseudolocal ones from scars \cite{Buca2023,Moudgalya2018,PhysRevB.101.165139,HoshoScars}. We also expect that the method could be advantageous in the simulation of NMR experiments where one needs to compute the Fourier transform of $\tr (Z_{tot}(t)Z_{tot}(0))$ \cite{Seetharam2023}. Indeed, extracting the dynamical symmetries solves the problem without having to time-evolve the system.

\vspace{0.5cm}

\begin{acknowledgments}
{\it Data availability -- } A minimal Julia code to reproduce our results is available at \url{https://github.com/nicolasloizeau/Opening-Krylov-space }.
\end{acknowledgments}
\vspace{0.5cm}

\begin{acknowledgments}
{\it Acknowledgments -- } We thank O. Castro Alvaredo, M. Fagotti, V. Mari\'c, L. Mazza, T. Prosen for useful discussions and feedback on the manuscript. N.L. and B.B were supported by a research grant (42085) from Villum Fonden.
This work was supported in part through the NYU IT High Performance Computing resources, services, and staff expertise. D.S. is grateful for ongoing support through the Flatiron Institute, a division of the Simons Foundation, and AFOSR through Grant FA9550-21-1-0236.
BB acknowledges funding by the French National Research Agency (ANR) under project ANR-24-CPJ1-0150-01.
\end{acknowledgments}

\bibliography{bib}

\begin{thebibliography}{140}
\bibitem [{foo({\natexlab{b}})}]{footnote1}%
  \BibitemOpen
  \href@noop {} {}\bibinfo {howpublished} {Recalling that our quenched state is $\rho(t)=e^{-\beta O_0(t)}/\tr(e^{- \beta O_0(t)})$ where $O_0(t)=e^{-iHt}O_0 e^{iHt}$ and the exponential can we expanded in power series. To first order, the dynamics of $\rho(t)$ is dictated by $O_0(t)$ and evolving $\rho$ is equivalent to evolving $O_0$. Therefore, studying the Krylov space generated by $O_0$ can be seen as studying a quench from Hamiltonian $O_0$ to first order} ({\natexlab{b}})\BibitemShut {NoStop}%
\end{thebibliography}

\end{document}


\title{Supplemental Material\\
Opening Krylov space to access all-time dynamics via dynamical symmetries
}
\author{Nicolas Loizeau}
\affiliation{Niels Bohr Institute, University of Copenhagen, Copenhagen, Denmark}
\author{Berislav Bu\v ca}
\affiliation{Niels Bohr Institute, University of Copenhagen, Copenhagen, Denmark}
\affiliation{Universit\'{e} Paris-Saclay, CNRS, LPTMS, 91405, Orsay, France}
\affiliation{Clarendon Laboratory, University of Oxford, Parks Road, Oxford OX1 3PU, United Kingdom} 
\author{Dries Sels}
\affiliation{Department of Physics, New York University, New York, NY, USA}
\affiliation{Center for Computational Quantum Physics, Flatiron Institute, New York, NY, USA}

\date{\today}
\maketitle

\newpage
\onecolumngrid

\def\theequation{S\arabic{equation}}
\setcounter{equation}{0}


\onecolumngrid

\section{Eigenmodes of the Liouvilian in ideal cases}
Here we give analytical solutions for the eigenmodes of the Liouvillian in the ideal cases $b_n=n$ and $b_n=\sqrt{n}$. The eigenvalue equation is the time-independent Schrödinger equation: 
\begin{align}
    \omega\varphi_n&=i\partial_t\varphi_n\\
    &=i\left(b_{n}\varphi_{n-1}-b_{n+1}\varphi_{n+1}\right)\label{eq:recurrence}
\end{align}
\subsubsection{Linear growth}
Let's check that when $b_n=n$, then 
\begin{align}
    \varphi_n(\omega) = \frac{i^n M_n(\omega)}{n!}
\end{align}
where $M$ is defined as $M_n(x)=i^{n}n!\sum_{k=0}^n 2^k \binom{n}{k}\binom{(-ix-1)/2}{k}$ and satisfies the recurrence relation $M_0(x)=1$, $M_1(x)=x$, $M_{n+1}(x)=xM_n(x)-n^2M_{n-1}(x)$ \cite{oeis_m}.
Start with the left hand side of \eqref{eq:recurrence} with $b_n=n$:
\begin{align}
    i\left(n\varphi_{n-1}-(n+1)\varphi_{n+1}\right)&=i\left(\frac{i^{n-1}n }{(n-1)!}M_{n-1}(\omega)-\frac{i^{n+1}(n+1)}{(n+1)!}M_{n+1}(\omega)\right)\\
    &=i^n\left(\frac{n^2}{n!}M_{n-1}(\omega)+\frac{1}{n!}M_{n+1}(\omega)\right)\\
    &=\omega\frac{i^n M_n(\omega)}{n!}=\omega\varphi_n(\omega)
\end{align}
where in the last step we used the recursion relation for $M$. The open boundary condition leads to the constraint on $\omega$:
\begin{equation}
    \left(\omega+2i(l+1)\right)M_l(\omega)=(2l+1)lM_{l-1}(\omega).
    \label{eq:linear_constraint}
\end{equation}

\subsubsection{Square root growth}
Here we check that if $b_n=\sqrt{n}$, then
\begin{align}
    \varphi_n = \frac{i^n\Tilde{H}_n(\omega)} {\sqrt{n!}}
\end{align}
where $\Tilde{H}_n=\frac{1}{2^{n/2}}H_n\left(\frac{x}{\sqrt{2}}\right)$ is the $n$-th modified Hermite polynomial and satisfies the recurrence relation $\Tilde{H}_0(x)=1$, $\Tilde{H}_1(x)=x$, $\Tilde{H}_n(x)=x\Tilde{H}_{n-1}(x)-(n-1)\Tilde{H}_{n-2}(x)$ \cite{oeis_h}. Again, let's start from the right side of \ref{eq:recurrence}:
\begin{align}
    i\left(\sqrt n\varphi_{n-1}-\sqrt {(n+1)}\varphi_{n+1}\right)&=i\left(\frac{i^{n-1}\sqrt{n}}{\sqrt{(n-1)!}}\Tilde{H}_{n-1}(\omega)- \frac{i^{n+1}\sqrt{n+1}}{\sqrt{(n+1)!}}\Tilde{H}_{n+1}(\omega)\right)\\
    &=i^n\left(\frac{n}{\sqrt{n!}}\Tilde{H}_{n-1}(\omega)+\frac{1}{\sqrt{n!}}\Tilde{H}_{n+1}(\omega)\right)\\
    &=\frac{i^n}{\sqrt{n!}}\omega \Tilde{H}_{n}(\omega)=\omega\varphi_n
\end{align}
where in the last step we used the recursion relation for $\Tilde{H}$. The open boundary condition leads to the constraint on $\omega$:
\begin{equation}
    \left(\omega+2i\sqrt{l+1}\right)\Tilde{H}_l(\omega)=\left(l+\sqrt{l(l+1)}\right)\Tilde{H}_{l-1}(\omega).
\end{equation}

\begin{figure}[h]
\centering
\includegraphics[width=1\textwidth]{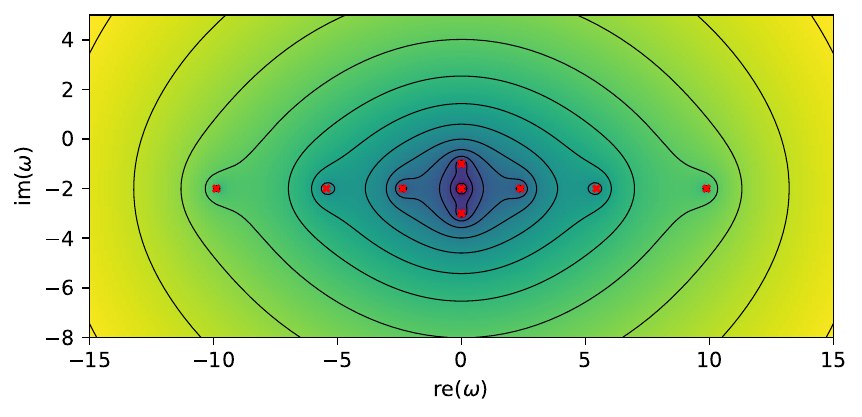}
\caption{To double check the constraint \eqref{eq:linear_constraint} we plot polynomial $P_l(\omega)=-\left(\omega+2i(l+1)\right)M_l(\omega)+(2l+1)lM_{l-1}(\omega)$ for $l=8$. The red crosses are the eigenvalues obtained by exact diagonalization of the Krylov matrix. They match with the roots of $P$.}
\end{figure}

\newpage
\section{Convergence}
\begin{figure}[h]
\centering
\includegraphics[width=0.8\textwidth]{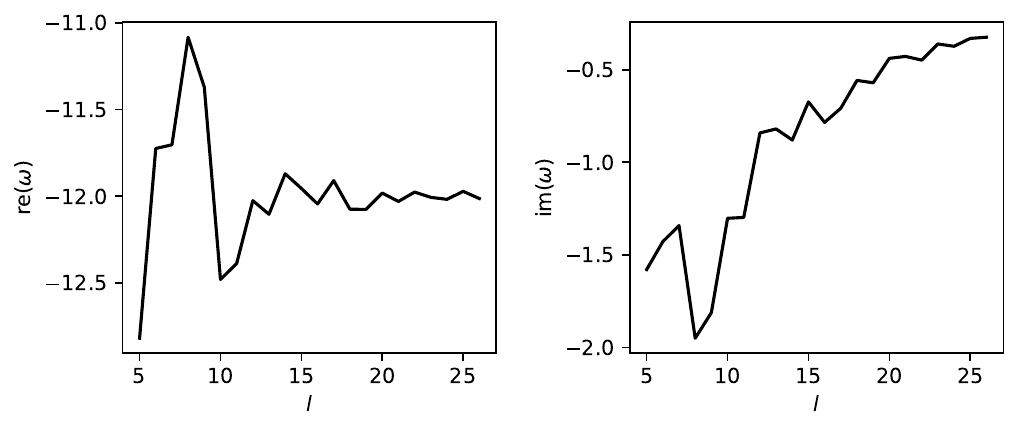}
\caption{Convergence of the perpetual dynamical symmetry in the XXZ model versus length of the Krylov chain.}
\label{fig:convergence}
\end{figure}

\begin{figure}[h]
\centering
\includegraphics[width=0.5\textwidth]{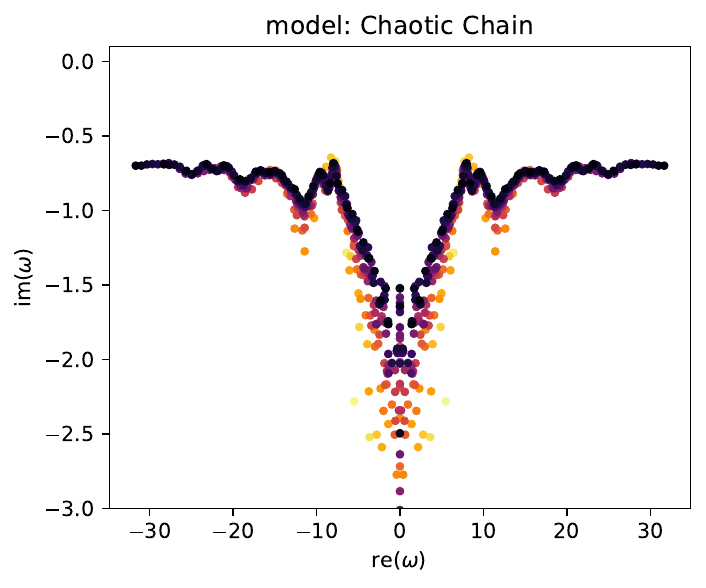}
\caption{Convergence of the spectrum of the open Krylov chain for the chaotic chain model. Each color correspond to a different truncation length. Darker is larger and darkest is $l=40$ sites.}
\label{fig:convergence_cc}
\end{figure}

We show how the method converges versus $l$ the length of the Krylov chain, i.e versus the position of the open boundary condition in figures \ref{fig:convergence} and \ref{fig:convergence_cc}.

\newpage
\section{Pauli truncation}
The main numerical limitation of the method is the difficulty to compute Lanczos coefficients $b_n$ for large $N$. In the case of the ZZX model, we can only exactly compute the first 26 coefficients. However, we obtain better results by also truncating the operators in the Pauli strings basis. This allows us to reach larger Lanczos step $n$. In ref \cite{paulistrings1} we have shown that we can compute long time dynamics of chaotic systems very efficiently by numerically integrating $\dot O(t)=i[H,O]$ and truncating $O(t)$ in the Pauli basis at each time step. This suggests that there is relevant information in the higher order Lanczos operators $O_n$ and Lanczos coefficients $b_n$ even if we cannot compute them exactly.
We propose the following truncated Lanczos algorithm :

\begin{algorithm}[H]
\caption{Pauli truncated Lanczos algorithm. At each step of the Lanczos algorithm, we truncate operator $O_n$ by only keeping $M$ Pauli strings with largest weight. }
\begin{algorithmic}
\State $O_0 \gets O_0/|O_0|$
\State $O_1\gets [H, O_0]$
\State $b\gets|O_1|$
\State $O_1\gets O_1/b$
\State $bs=[b]$

\For{$ 1 \leq n \leq l$ }
\State $O_2\gets [H, O_1]-b\cdot O_0$
\State $O_2\gets$ truncate($O_2, M$)
\State $b \gets |O_2|$
\State $O_2 \gets O_2/b$
\State $O_0 \gets O_1$
\State $O_1 \gets O_2$
\State push $b$ to $bs$

\EndFor

\State \Return {$bs$}
\end{algorithmic}
\end{algorithm}
This algorithm is used to compute the orange line on figure 4a of the main paper, and we show the Liouvillian spectrum obtained with this method on the figure below.

\begin{figure}[h]
\centering
\includegraphics[width=1\textwidth]{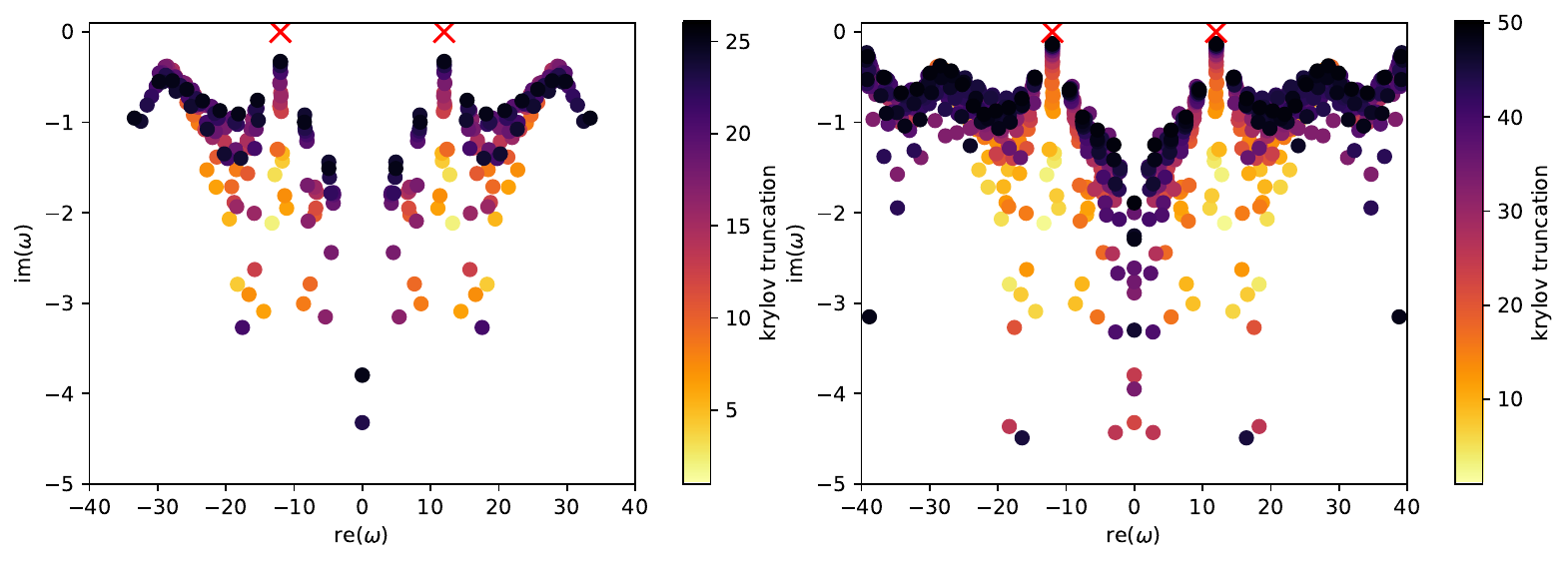}
\caption{Convergence of the spectrum of the open Krylov chain for the XXZ model. Each color correspond to a different truncation length. Darker is larger. On the left plot, the largest truncation is $n=26$. For this value of $n$, the $b_n$ are computed exactly. On the left plot, the largest truncation is $n=50$. We can reach Krylov truncation length larger than $26$ by also truncating the operators in the Pauli basis. Here, we only keep the $2^{22}$ strings with the largest weight. This truncation in the Pauli basis allows us to reach better convergence of the dynamical symmetry.}
\label{fig:truncated_lanczos}
\end{figure}

\newpage

\section{Recursive algorithm}
Here we suggest a more refined approach to extract unknown dynamical symmetries.
Assume we have a Hamiltonian $H$ and are interested of constructing a dynamical symmetry with frequency $\omega$. 
In order to achieve better results without knowing anything about the structure of the dynamical symmetry, we propose the following procedure:
\begin{enumerate}
    \item Generate and random local operator $O_0$.
    \item Diagonalize the Liouvilian using Lanczos algorithm starting from seed $O_0$, and select the eigenmode $A_\omega$ with frequency closest to $\omega$.
    \item Set $O_0 \gets \text{truncate}(A_\omega)$ and repeat 2. until desired convergence.
\end{enumerate}
where `truncate' project $A_\omega$ on the $k$-local subspace and $k$ has to be chosen empirically depending on the problem. Truncating $A_\omega$ ensures that the dynamical symmetry is local and makes the problem numerically tractable.

\begin{figure}[h]
\centering
\includegraphics[width=0.8\textwidth]{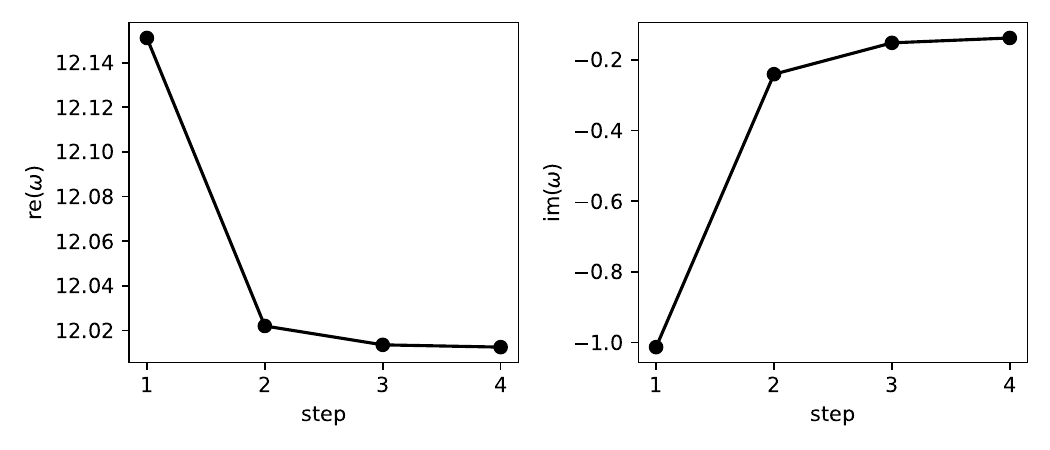}
\label{fig:truncated_lanczos}
\caption{Convergence of the recursive algorithm on the ZZX model. The procedure is initialized with a random 3-local operator and $A_\omega$ is truncated at each step by only keeping its 3-local part.}
\end{figure}

\newpage

\section{Comparison to Terentenkov et al}
\begin{figure}[h]
\centering
\includegraphics[width=0.8\textwidth]{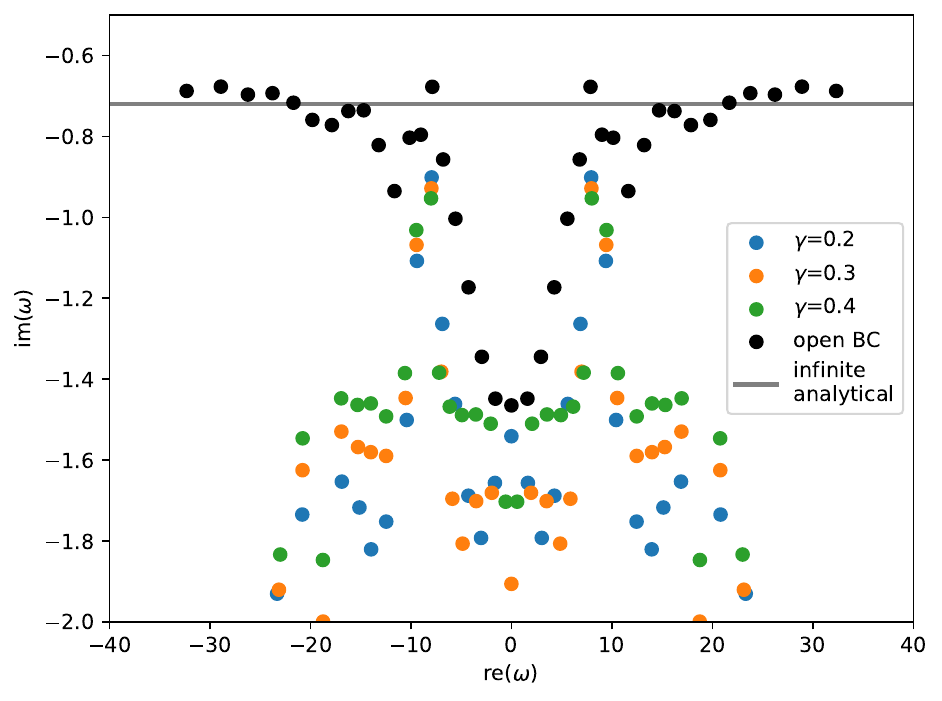}
\caption{Comparison of the Liouvillian spectrum of the chaotic chain model extracted with Terentekov and al's method (color) and our method (black).}
\label{fig:teretenkov1}
\end{figure}

\begin{figure}[h]
\centering
\includegraphics[width=0.8\textwidth]{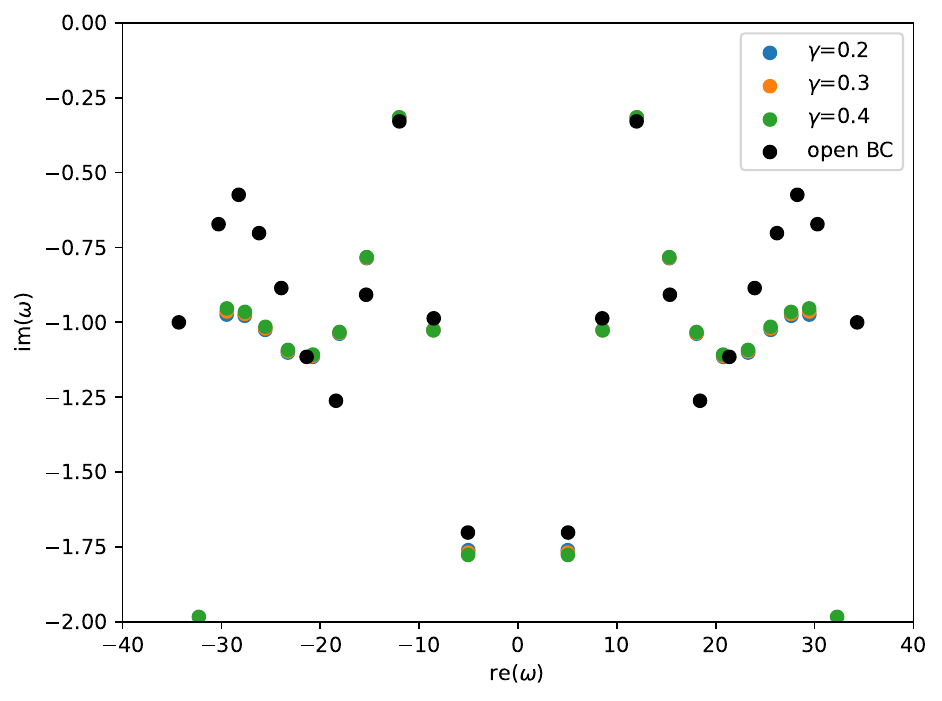}
\caption{Comparison of the Liouvillian spectrum of the XXZ model extracted with Terentekov and al's method (color) and our method (black).}
\label{fig:teretenkov2}
\end{figure}

In ref.~\cite{Teretenkov2024}, a similar method is described. Instead of adding an open boundary condition to the Krylov chain, the coefficients are extrapolated linearly and dissipation is added to the chain. In this model, the Krylov chain equation is
\begin{align}
    \partial_t \varphi_n=b_{n}\varphi_{n-1}-b_{n+1}\varphi_{n+1} +i a_n\varphi_{n}
\end{align}
with 
\begin{equation}
    a_n=
\begin{cases}
 0,& n \leq n_{\rm max}\\
 i\gamma(n- n_{\rm max}),              & n\geq n_{\rm max}+1
\end{cases}
\end{equation}
In figures \ref{fig:teretenkov1} and \ref{fig:teretenkov2} we compare our results to this method. Both methods yield quite different results. In the chaotic chain model, Terentekov and al's method does not produce eigenmodes that accumulate around the infinite size analytical result, while our method does.
In the XXZ model, both methods reproduce the local dynamical symmetry at $\omega=12$ with similar precision. However, the two methods do not agree for the non-local modes. 

\section{Master equation for the boundary condition}

The exact time evolution equation for operator $O$ is $\dot O = i[H, O]$.
The hard boundary condition at site $L$ of the Krylov chain corresponds to forbidding $O$ to ever reach $O_L$. This can be achieved by removing the part of $\dot O$ that overlaps with $O_L$:
\begin{align}
    \dot O = i[H, O]-\frac{1}{2^N}\tr\left( O_L^\dagger i [H,O]\right)O_L
\end{align}

Now, in Krylov space we have
\begin{align}
    \dot O(t) &= \frac{1}{2^N}\sum_1^{L}i^n \varphi'_n(t)O_n\\
    &=\frac{1}{2^N}\sum_1^{L-1}i^n \varphi'_n(t)O_n + \frac{1}{2^N}i^L \varphi'_L(t)O_L
\end{align}
and let's plug the equation for the open boundary condition and use $\tr(O O_n)=\frac{i^n}{2^N} \varphi(t)$
\begin{align}
    \dot O(t)&=\frac{1}{2^N}\sum_1^{L-1}i^n \varphi'_n(t)O_n + \frac{1}{2^N}i^L \left((b_L+b_{L+1})\varphi_{L-1}-2b_{L+1}\varphi_L\right)O_L\\
    &=\frac{1}{2^N}\sum_1^{L-1}i^n \varphi'_n(t)O_n + i b_{L+1} \tr( O_{L-1}^\dagger O)O_L -2b_{L+1} \tr(O_L^\dagger O)O_L\\
    &=i[H, O]-\frac{1}{2^N}\tr\left( O_L^\dagger i [H,O]\right)O_L + i b_{L+1} \tr( O_{L-1}^\dagger O)O_L -2b_{L+1} \tr(O_L^\dagger O)O_L
\end{align}

\section{Additional figures}

\begin{figure*}[h]
\centering
\includegraphics[width=0.9\textwidth]{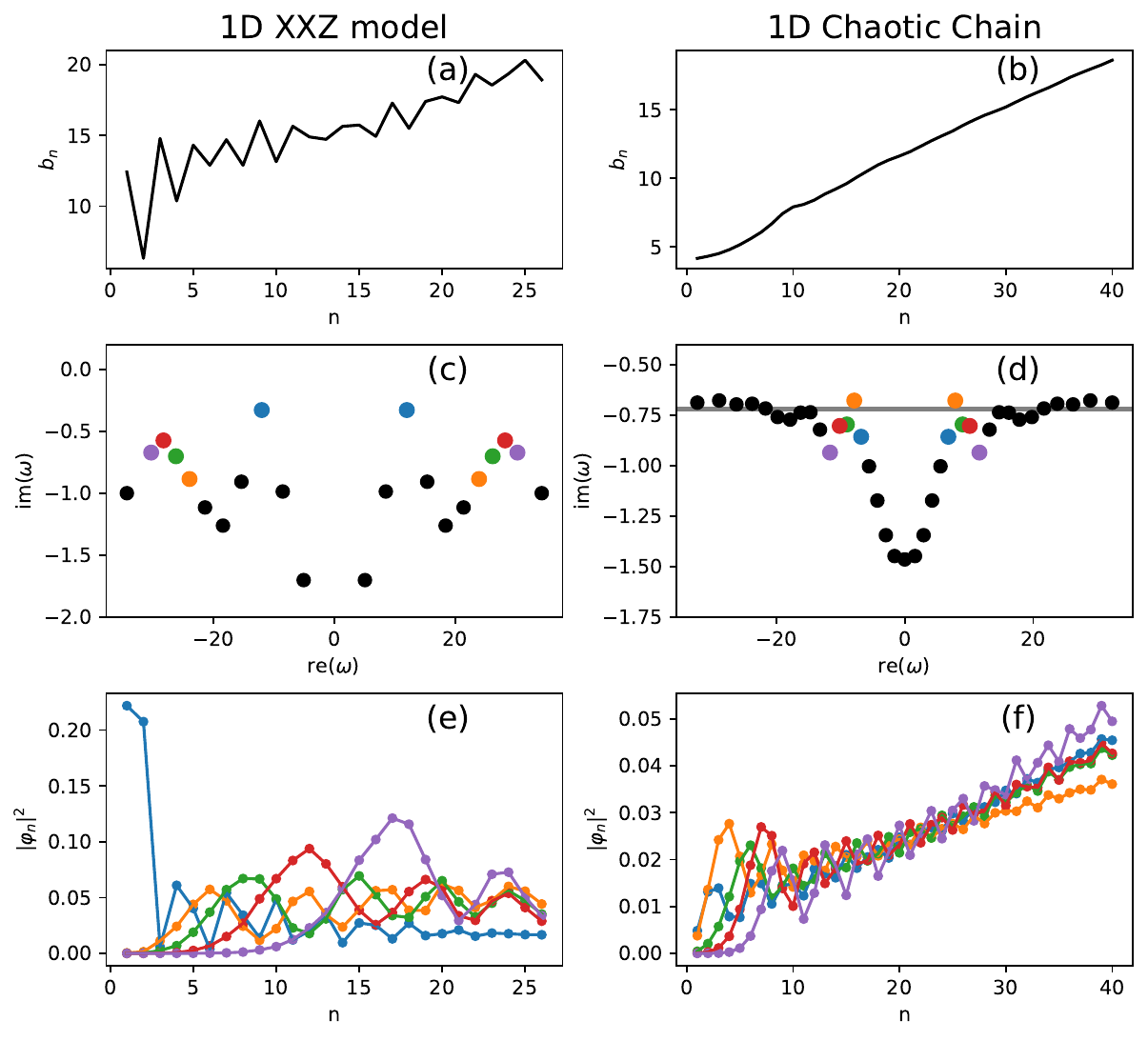}
\caption{
(\textbf{a, b}) Lanczos coefficients of the $XXZ$ model and chaotic chain.
(\textbf{c, d}) Spectrum of the open Krylov chain for the $XXZ$ model and chaotic chain. In the XXZ model, we recover the dynamical symmetry at $\textup{re}(\omega)=12$ (blue). In the chaotic chain, the grey horizontal line indicates $\textup{im}(\omega)=-2\lambda$ where $\lambda$ is the growth rate of the Lanczos coefficients shown in (B). The inset on the left plot shows exact results in the ideal case $b_n=n$. Eigenvectors corresponding to the color markers are displayed below. 
(\textbf{e, f}) Eigenvectors of the Krylov chain expressed in the Krylov basis. The dynamical symmetry at $\textup{re}(\omega)=12$ (blue) in the XXZ model is localized on the left part of the chain, meaning that is is also spatially local. For simplicity we only show the eigenvectors that correspond to the eigenvalues in colors in (c,d).  
}
\label{fig:main_sup}
\end{figure*}

\bibliography{bib}